# Deterministic patterned growth of high-mobility large-crystal graphene: a path towards wafer scale integration

Vaidotas Mišeikis[1,2,3], Federica Bianco[4], Vittorio Pellegrini[2], Marco Romagnoli[3], Camilla Coletti[1,2]

[1] Center for Nanotechnology Innovation @ NEST, Istituto Italiano di Tecnologia, P.za S. Silvestro 12, 56127 Pisa, Italy

[2] Graphene Labs, Istituto Italiano di Tecnologia, Via Morego 30, 16163 Genova, Italy

[3] Consorzio Nazionale Interuniversitario per le Telecomunicazioni (CNIT), Via Moruzzi 1, 56124 Pisa, Italy

[4] NEST, Istituto Nanoscienze-CNR and Scuola Normale Superiore, P.za S. Silvestro 12, 56127 Pisa, Italy

We demonstrate rapid deterministic (seeded) growth of large single-crystals of graphene by chemical vapour deposition (CVD) utilising pre-patterned copper substrates with chromium nucleation sites. Arrays of graphene single-crystals as large as several hundred microns are grown with a periodicity of up to 1 mm. The graphene is transferred to target substrates using aligned and contamination-free semi-dry transfer. The high quality of the synthesised graphene is confirmed by Raman spectroscopy and transport measurements, demonstrating room-temperature carrier mobility of 21 000 $cm^2$ / V s when transferred on top of hexagonal boron nitride. By tailoring the nucleation of large single-crystals according to the desired device geometry, it will be possible to produce complex device architectures based on single-crystal graphene, thus paving the way to the adoption of CVD graphene in wafer-scale fabrication.

## Introduction

CVD synthesis of graphene on catalytically-active substrates has emerged as the most promising approach for large-area production of graphene [1]. The self-limiting nature of CVD growth on metals such as copper (Cu) and platinum allows synthesis of large-scale homogeneous films of monolayer graphene. However, electrical characterization of polycrystalline samples of CVD graphene reveals that the presence of grain boundaries causes significant degradation of the electric performance, compared to pristine material obtained by mechanical exfoliation of flakes [2]. As demonstrated initially by Petrone et al [3], samples fabricated using single-crystals of CVD graphene can have electrical performance comparable to that of exfoliated flakes [4]. Furthermore, recent reports have shown that by fully encapsulating CVD graphene with suitable materials such as hexagonal boron nitride (h-BN), low-temperature charge carrier mobility above 300 000 $cm^2$ / V s [5] or even 3 000 000 $cm^2$ / V s [6] can be achieved.

Over the last few years the synthesis of large-crystal graphene has attracted a huge scientific interest, with significant advances in the achievable crystal size [7–10]. Recent work has reported single-crystals of graphene measuring 1 cm [10] and, using copper/nickel alloy as the growth substrate, even 4 cm [11]. Inevitably, these approaches still produce randomly-distributed crystals of graphene, which limits their applicability to scaled production of graphene devices. Furthermore, the commonly-used transfer methods either allow scalability while introducing significant performance degradation, or are limited to transferring areas of several tens of $\mu m^2$ [5].

For many applications the size of individual graphene devices is limited to tens or hundreds of microns, easily achievable by the current methods of single-crystal synthesis, however, random spatial distribution of graphene crystals in such samples makes polycrystalline graphene preferable for wafer-scale integration. This issue could be mitigated by selectively pre-determining the nucleation sites for graphene crystals according to the target architecture, which could allow the fabrication of large and complex circuits utilising completely monocrystalline graphene. Patterned growth using polymer-based nucleation seeds was first reported by Wu et al [12], however, only high-density arrays of 10-20 µm crystals were demonstrated. Arrays of similar dimensions were recently presented by Song et al, using poly(methyl methacrylate) (PMMA) seeds to nucleate graphene on top of CVD-grown h-BN [13].

In this work we present a method to selectively pattern the Cu growth substrate using chromium (Cr) nucleation seeds, which allows deterministic nucleation of large-crystal graphene, measuring several hundred microns. The nucleation density is highly-controlled by the combined use of natively oxidised Cu foils, non-reducing annealing and sample enclosure [14], and measuring as low as 10 crystals per $mm^2$. We also demonstrate a clean semi-dry transfer procedure allowing aligned placement of graphene crystals on target substrates. Figure 1 (a-d) shows the proposed process flow for scalable fabrication of single-crystal devices. Initially, the device layout is designed (a), followed by the deposition of Cr seeds on copper substrate (b). Large crystals of graphene are grown in a deterministic manner (c) and are then transferred to the target substrate, followed by conventional sample fabrication (d). Combining deterministic growth of large graphene crystals and aligned transfer allows the fabrication of fully-tailored samples utilising high-quality single-crystal graphene. In principle, this technique is scalable to the size of full wafers, paving the way for wafer-scale integration of high-quality graphene.

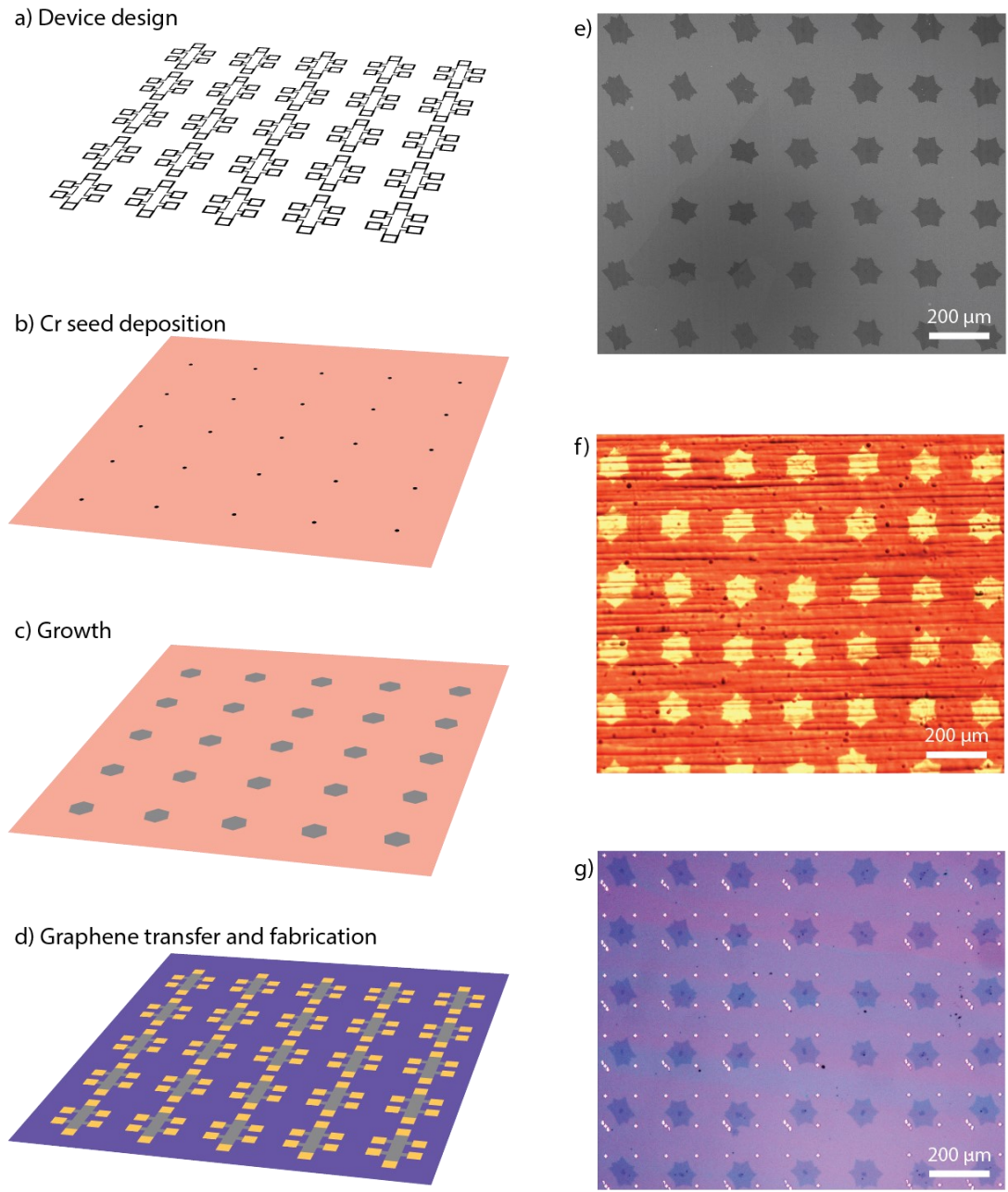

**Figure 1.** a-d) Proposed process flow for deterministic growth of single-crystal graphene. e) SEM image of an array of graphene single-crystals. f) Optical image of a graphene array on oxidised Cu foil. g) Optical image of single-crystal array transferred on Si/SiO$_2$ substrate with alignment markers. In figures (e-g) the array periodicity is 200 µm and crystal size is ~100 µm.

### Experimental

The initial step of seeded growth of large-crystal graphene is the preparation and patterning of the growth substrate. Natively-oxidised copper foils (Alfa-aesar, 99.8%, 13382) were used in this work, which were electropolished [14] in order to clean the surface and to improve the surface morphology. Nucleation sites were then patterned on the foil by optical lithography and thermal evaporation of 25 nm of chromium, followed by lift-off. The growth of large single-crystals was performed using a low-pressure (25 mbar) CVD growth procedure using a commercially-available Aixtron BM Pro cold-wall reactor. As described in our previous work [14], the substrates were annealed for 10 minutes in non-reducing argon atmosphere in order to preserve surface oxidation. Graphene was grown at a temperature of 1060 °C with a gas flow comprising 900 standard cubic centimetres per minute (sccm) of argon, 100 sccm of hydrogen (H$_2$) and 1 sccm of methane (CH$_4$).

To minimise the transfer-related contamination and to allow precise placement of graphene arrays on the target substrates, we developed a semi-dry aligned transfer technique which can be conveniently scaled up. A thin PMMA carrier membrane was spin-coated on the Cu foil with seeded graphene and the graphene was detached from the growth substrate using electrochemical delamination [15,16]. As shown in the inset to Fig. 3(a), a semi-rigid Kapton frame was used to handle a suspended PMMA membrane carrying the seeded array of graphene crystals, which was rinsed in deionised water, dried, and aligned to the target substrates using a micro-mechanical stage. The target substrates were mildly heated to improve the adhesion of graphene, and the PMMA was finally removed in acetone and isopropanol. More details about this transfer technique can be found in the Supplementary Information.

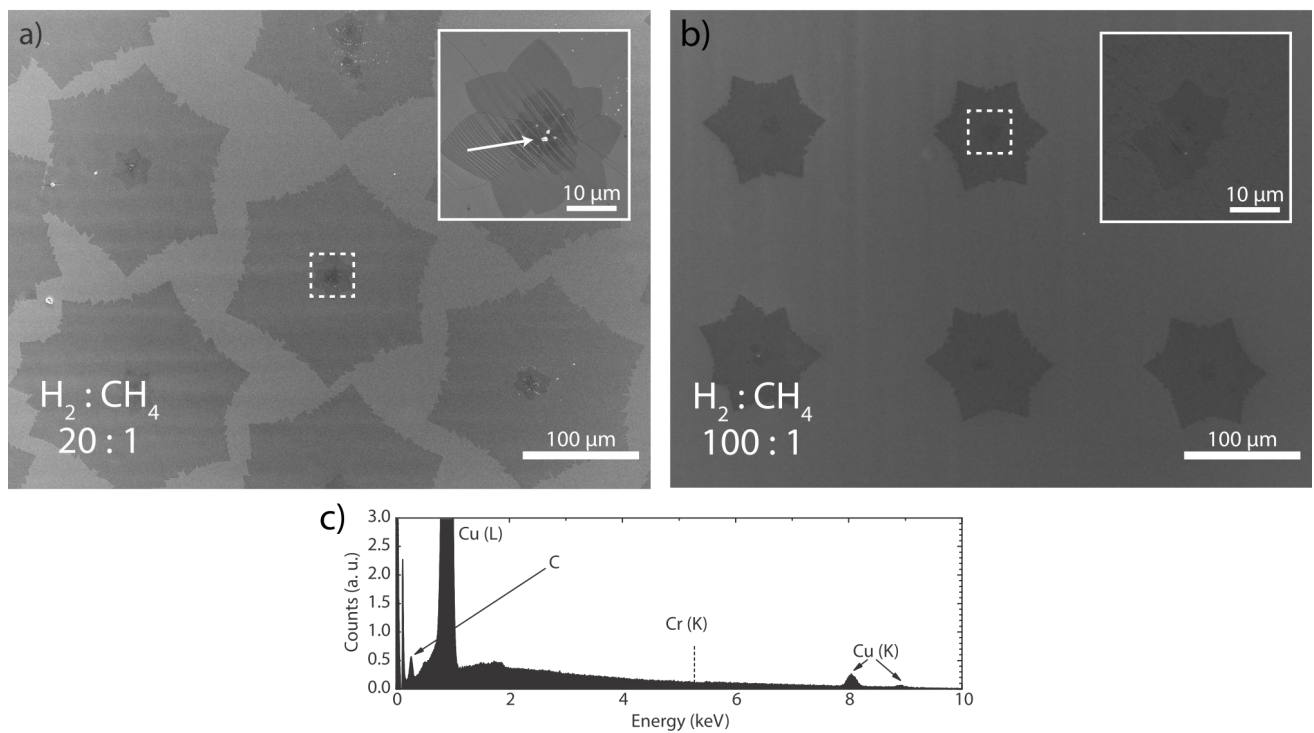

**Figure 2.** Comparison of SEM images of graphene arrays synthesised using low (a) and high (b) ratio of hydrogen to methane. Insets show magnified SEM images obtained from the centers of typical crystals, as indicated by the dashed lines. c) EDX spectrum obtained from the center of a crystal synthesised using high flow of hydrogen.

Raman spectroscopy was performed to study the quality and homogeneity of the crystals which were transferred via aligned semi-dry transfer on top of exfoliated h-BN flakes (HQ Graphene). A Renishaw InVia system was used, equipped with a 532 nm laser and 100x objective lens, providing a spot size of ~1 μm.

To investigate the charge carrier transport characteristics of seeded large-crystal graphene, Hall bar devices were fabricated on top of h-BN flakes by electron-beam lithography and reactive ion etching. The devices were studied at room temperature (RT) by measuring the electric-field effect using a 4-terminal configuration. The measurements were performed using a lock-in technique, by passing a 1 nA current between the source-drain contacts of the Hall bar and measuring the voltage drop $V_{xx}$ along the side contacts as a function of applied back-gate voltage.

**Results and discussion**
Panels (e,f) of figure 1 show a representative square array of large-crystal graphene produced using patterned nucleation and panel 1(g) displays the aligned transfer of this array to a pre-patterned wafer containing a marker grid. Clearly, at the graphene growth temperatures, the deposited Cr thin films form particles which act as effective nucleation sites for graphene, thus enabling growth of graphene patterns. The representative crystals displayed in panels (e-g) have an average diameter of 100 μm and a periodicity of 200 μm, which is an order of magnitude larger than previous reports [12,13]. As shown in the supplementary figure S3(a), such arrays present an extremely well-controlled nucleation over a large area, with non-seeded crystals comprising less than 10% of the total number. The high control on spurious nucleation is possible thanks to the fact that growth is carried on using oxidised Cu foils and a non-reducing atmosphere, as reported in [14]. It is relevant to mention that by simply varying the substrate patterning, it is possible to easily produce arrays with different periodicity (up to 1 mm), crystal size (up to 350 μm) and pattern (e.g., hexagonal, square), as shown in the Supplementary Information. Due to the short annealing and fast growth rate, the total synthesis time of arrays containing 350 μm crystals is notably short – i.e., around 2 hours including the heating and cooling of the CVD system.

As demonstrated in Fig. 2, the ratio of hydrogen to methane is found to be an important parameter for regulating the growth rate of graphene as well as controlling the presence of chromium after the growth. During preliminary experiments we found that although lower $H_2:CH_4$ ratio (20:1) provided fast growth rates (~15 μm/min), the synthesised samples contained parasitic particles of Cr at the centre of each crystal (inset to Fig. 2(a), particle indicated by an arrow). Increasing this ratio to 100:1 caused a slight reduction to the growth rate (~12 μm/min). However, the higher hydrogen flow had an advantageous effect of removing the chromium particles. The patterned nucleation evident in these samples (Fig. 2(b)) clearly indicates that during the initial stages of the growth, Cr particles were present on the surface, successfully seeding the graphene crystals, and were then gradually removed during the growth process. The lack of chromium on the surface of the foil was confirmed using high-magnification SEM imaging (inset to Fig. 2(b)) as well as energy-dispersive X-ray spectroscopy. Figure 2(c) shows an EDX spectrum obtained at an accelerating voltage of 10 kV at the centre of a seeded graphene crystal grown using a high $H_2:CH_4$ ratio. The spectrum is clearly dominated by the L- and K-lines of Cu substrate. Additionally, the K-line of carbon can be seen at 0.28 keV. No evidence of L-line of Cr (0.57 keV) or K-line of Cr (5.41 keV) was observed.

It is well-known that the major limiting factor for the quality of graphene samples is the substrate [17,18] as well as the contam-ination and damage caused during the transfer procedure [19]. To mitigate the negative effects of low-quality substrates and to focus on the intrinsic quality of the synthesised graphene, we used aligned semi-dry transfer to place the graphene on top of exfoliated flakes of h-BN, which is known as the optimal substrate for graphene-based devices [20]. Figure 3(a) shows a seeded crystal transferred on top of an h-BN flake with a thickness of 30 nm. With increasing scientific effort and recent progress in CVD synthesis of wafer-scale h-BN [21], this material has high potential to become widely-adopted for large-scale graphene.

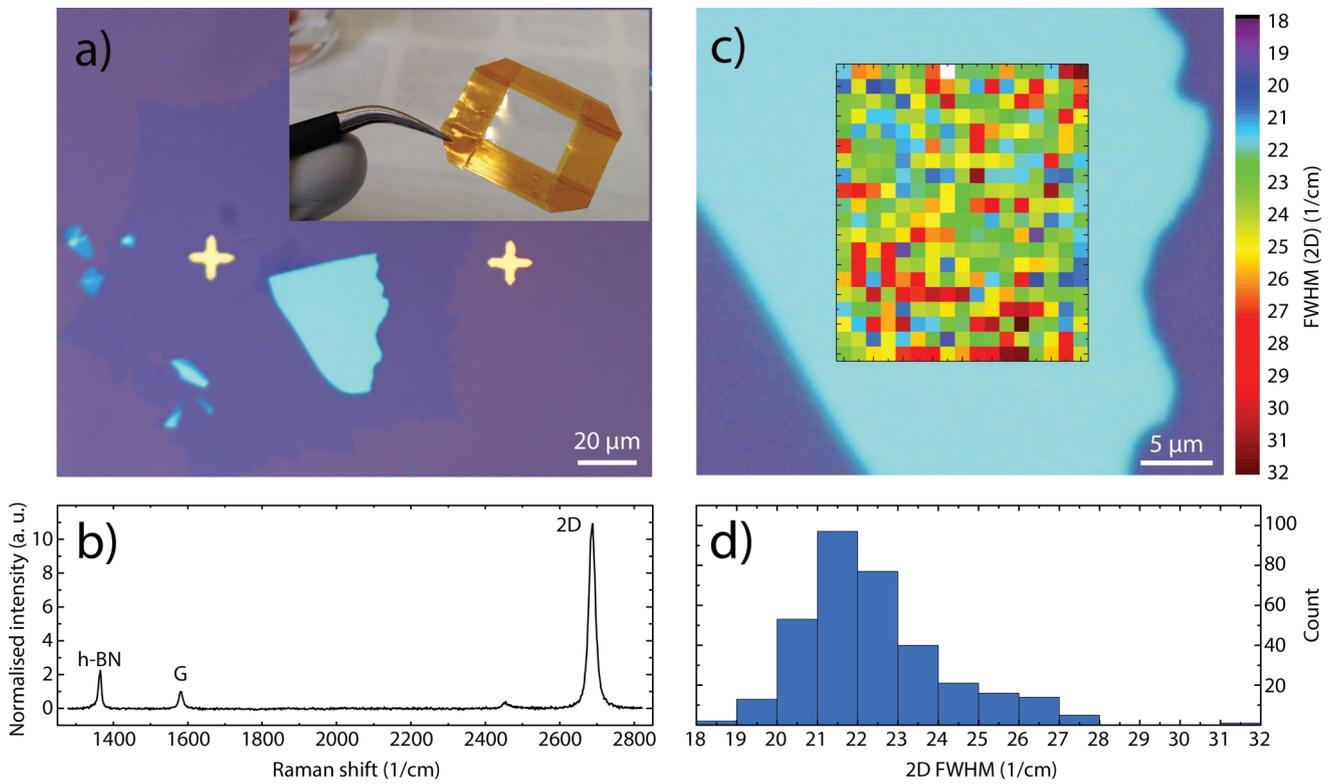

**Figure 3.** Seeded single-crystal graphene on exfoliated h-BN (colored in light blue). a) Optical image of a seeded single-crystal placed on top of a flake of h-BN using aligned transfer. Inset: A photo of PMMA carrier membrane with graphene crystals, suspended over a Kapton frame. b) A representative graphene Raman spectrum obtained on h-BN. c) Raman map of the width of 2D peak. d) Histogram of the $\Gamma_{2D}$ values obtained by mapping.

Figure 3(b) shows a typical room-temperature Raman spectrum obtained from the seeded graphene transferred on top of an h-BN flake. The spectrum has 3 prominent features: the characteristic h-BN peak at 1364 cm$^{-1}$, along with the two peaks commonly observed in high-quality graphene, the G-peak at 1582 cm$^{-1}$ and the 2D peak at 2687 cm$^{-1}$. No disorder-related D peak at 1350 cm$^{-1}$ was observed, indicating an extremely low density of lattice defects. The D-peak was also negligible at the centre of the crystals, further confirming the lack of negative effects from the nucleation seeds (Raman map presented in Supplementary Information). The intensity ratio of 2D- and G-peaks is in excess of 10 to 1. In addition, a remarkably low width of the 2D-peak $\Gamma_{2D}$ was measured at 21 cm$^{-1}$. The width of this peak is a reliable indicator of the sample quality, as it is highly sensitive to strain variation which is a known source of charge scattering in high-quality graphene [22,23]. Therefore, the variation of $\Gamma_{2D}$ over large areas of the sample was investigated via spatially-resolved Raman mapping. Figure 3(c) shows a map of the $\Gamma_{2D}$ obtained over an area of 17x20 µm$^2$, whereas Fig. 3(d) shows the histogram of $\Gamma_{2D}$ values obtained in the area mapped, with the majority of $\Gamma_{2D}$ values obtained on h-BN being within 20-23 cm$^{-1}$. This indicates the high structural quality of graphene with small strain variations [22,23]. Lower values of $\Gamma_{2D}$ can typically be observed only in graphene samples with full (double-sided) h-BN encapsulation [5].

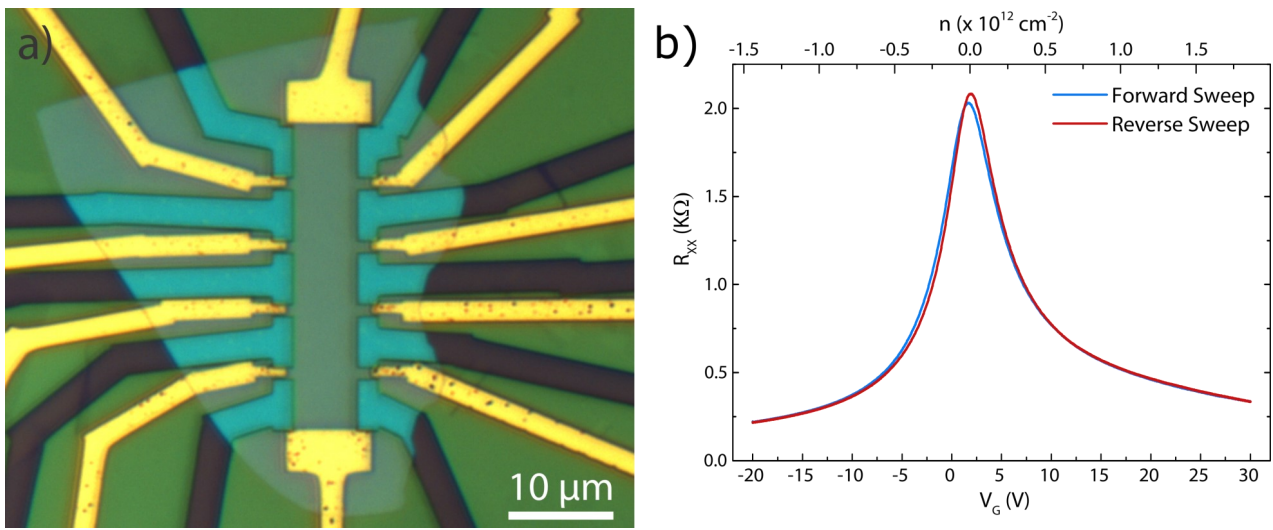

**Figure 4.** Hall Bar device for transport measurements. a) Optical image of the Hall bar device fabricated on h-BN for mobility measurements prior to the removal of PMMA etching mask. b) Room-temperature electric field effect of seeded graphene on top of h-BN as a function of applied gate voltage (bottom axis) and gate-induced carrier density (top axis).

Fig. 4 (a) is an optical micrograph of a typical Hall bar device fabricated on top of exfoliated hBN flakes and used for transport measurements. Figure 4(b) reports the measured RT electric field effect as a function of applied gate voltage (bottom axis) and gate-induced carrier density (top axis). A sharp and narrow resistivity peak is measured at 1.8 V, indicating a low residual carrier density in the sample, around 1.4 x 10$^{11}$ cm$^{-2}$. Only negligible hysteresis was observed when sweeping the gate voltage in both directions. By fitting the field effect data using the model reported by Kim *et al* [24], hole mobility are estimated to be in excess of 21 000 cm$^2$ / V s, whereas electron mobility are found to be about 13 500 cm$^2$ / V s. These values are comparable to the RT mobility reported on high-quality non-encapsulated graphene [3,11,20].

## Conclusion

To summarise, we have presented a method for fast and highly-controllable seeded growth of large single-crystal graphene, which can be deterministically transferred to target substrates. The periodicity and crystal size of the arrays can be flexibly varied with maximum values peaking at 1 mm and 350 µm, respectively. Both the synthetic method and the transfer approach presented in this work are scalable and lead to high quality graphene as confirmed by spectroscopic and transport measurements. Indeed, when transferred on top of h-BN, seeded graphene crystals present an ultra-low width of the Raman 2D peak $\Gamma_{2D}$ = 20-23 cm$^{-1}$ and room temperature hole mobility above 21 000 cm$^2$/V s. The quality of seeded single-crystal graphene is comparable to that of the best published results obtained on CVD graphene and exfoliated flakes. Taking into account the possibility of tailored synthesis of this material according to the target device architecture and the possibility for contamination-free and aligned transfer, this work provides a significant advance for adoption of CVD graphene in wafer-scale fabrication.

## Acknowledgments

We thank Stefano Roddaro of CNR@NEST and Alessandro Tredicucci of University of Pisa for fruitful discussions. This project has received funding from the European Union's Horizon 2020 research and innovation programme under grant agreement No. 696656-GrapheneCore1.

## References

(1) Li, X.; Cai, W.; An, J.; Kim, S.; Nah, J.; Yang, D.; Piner, R.; Velamakanni, A.; Jung, I.; Tutuc, E.; et al. Large-Area Synthesis of High-Quality and Uniform Graphene Films on Copper Foils. *Science* **2009**, *324* (5932), 1312–1314.

(2) Yu, Q.; Jauregui, L. a; Wu, W.; Colby, R.; Tian, J.; Su, Z.; Cao, H.; Liu, Z.; Pandey, D.; Wei, D.; et al. Control and Characterization of Individual Grains and Grain Boundaries in Graphene Grown by Chemical Vapour Deposition. *Nat. Mater.* **2011**, *10* (6), 443–449.

(3) Petrone, N.; Dean, C. R.; Meric, I.; van der Zande, A. M.; Huang, P. Y.; Wang, L.; Muller, D.; Shepard, K. L.; Hone, J. Chemical Vapor Deposition-Derived Graphene with Electrical Performance of Exfoliated Graphene. *Nano Lett.* **2012**, *12* (6), 2751–2756.

(4) Xiang, S.; Miseikis, V.; Planat, L.; Guiducci, S.; Roddaro, S.; Coletti, C.; Beltram, F.; Heun, S. Low-Temperature Quantum Transport in CVD-Grown Single Crystal Graphene. *Nano Res.* **2016**, No. Cvd, 1–8.

(5) Banszerus, L.; Schmitz, M.; Engels, S.; Dauber, J.; Oellers, M.; Haupt, F.; Watanabe, K.; Taniguchi, T.; Beschoten, B.; Stampfer, C. Ultrahigh-Mobility Graphene Devices from Chemical Vapor Deposition on Reusable Copper. *Sci. Adv.* **2015**, *1* (6), e1500222–e1500222.

(6) Banszerus, L.; Schmitz, M.; Engels, S.; Goldsche, M.; Watanabe, K.; Taniguchi, T.; Beschoten, B.; Stampfer, C. Ballistic Transport Exceeding 28 Mm in CVD Grown Graphene. *Nano Lett.* **2016**, *16* (2), 1387–1391.

(7) Li, X.; Magnuson, C. W.; Venugopal, A.; Tromp, R. M.; Hannon, J. B.; Vogel, E. M.; Colombo, L.; Ruoff, R. S. Large-Area Graphene Single Crystals Grown by Low-Pressure Chemical Vapor Deposition of Methane on Copper. *J. Am. Chem. Soc.* **2011**, *133* (9), 2816–2819.

(8) Wang, H.; Wang, G.; Bao, P.; Yang, S.; Zhu, W.; Xie, X.; Zhang, W.-J. Controllable Synthesis of Submillimeter Single-Crystal Monolayer Graphene Domains on Copper Foils by Suppressing Nucleation. *J. Am. Chem. Soc.* **2012**, *134* (8), 3627–3630.

(9) Zhou, H.; Yu, W. J.; Liu, L.; Cheng, R.; Chen, Y.; Huang, X.; Liu, Y.; Wang, Y.; Huang, Y.; Duan, X. Chemical Vapour Deposition Growth of Large Single Crystals of Monolayer and Bilayer Graphene. *Nat. Commun.* **2013**, *4*, 2096.

(10) Hao, Y.; Bharathi, M. S.; Wang, L.; Liu, Y.; Chen, H.; Nie, S.; Wang, X.; Chou, H.; Tan, C.; Fallahazad, B.; et al. The Role of Surface Oxygen in the Growth of Large Single-Crystal Graphene on Copper. *Science* **2013**, *342* (6159), 720–723.

(11) Wu, T.; Zhang, X.; Yuan, Q.; Xue, J.; Lu, G.; Liu, Z.; Wang, H.; Wang, H.; Ding, F.; Yu, Q.; et al. Fast Growth of Inch-Sized Single-Crystalline Graphene from a Controlled Single Nucleus on Cu-Ni Alloys. *Nat. Mater.* **2015**, *15* (1), 43–47.

(12) Wu, W.; Jauregui, L. A.; Su, Z.; Liu, Z.; Bao, J.; Chen, Y. P.; Yu, Q. Growth of Single Crystal Graphene Arrays by Locally Controlling Nucleation on Polycrystalline Cu Using Chemical Vapor Deposition. *Adv. Mater.* **2011**, *23* (42), 4898–4903.

(13) Song, X.; Gao, T.; Nie, Y.; Zhuang, J.; Sun, J.; Ma, D.; Shi, J.; Lin, Y.; Ding, F.; Zhang, Y.; et al. Seed-Assisted Growth of Single-Crystalline Patterned Graphene Domains on Hexagonal Boron Nitride by Chemical Vapor Deposition. *Nano Lett.* **2016**, *16* (10), 6109–6116.

(14) Miseikis, V.; Convertino, D.; Mishra, N.; Gemmi, M.; Mashoff, T.; Heun, S.; Haghighian, N.; Bisio, F.; Canepa, M.; Piazza, V.; et al. Rapid CVD Growth of Millimetre-Sized Single Crystal Graphene Using a Cold-Wall Reactor. *2D Mater.* **2015**, *2* (1), 14006.

(15) Wang, Y.; Zheng, Y.; Xu, X.; Dubuisson, E.; Bao, Q.; Lu, J.; Loh, K. P. Electrochemical Delamination of CVD-Grown Graphene Film: Toward the Recyclable Use of Copper Catalyst. *ACS Nano* **2011**, *5* (12), 9927–9933.

(16) Gao, L.; Ren, W.; Xu, H.; Jin, L.; Wang, Z.; Ma, T.; Ma, L.-P.; Zhang, Z.; Fu, Q.; Peng, L.-M.; et al. Repeated Growth and Bubbling Transfer of Graphene with Millimetre-Size Single-Crystal Grains Using Platinum. *Nat. Commun.* **2012**, *3*, 699.

(17) Chen, J.; Jang, C.; Xiao, S.; Ishigami, M.; Fuhrer, M. S. Intrinsic and Extrinsic Performance Limits of Graphene Devices on SiO2. *Nat. Nanotechnol.* **2008**, *3* (4), 206–209.

(18) Bolotin, K. I.; Sikes, K. J.; Jiang, Z.; Klima, M.; Fudenberg, G.; Hone, J.; Kim, P.; Stormer, H. L. Ultrahigh Electron Mobility in Suspended Graphene. *Solid State Commun.* **2008**, *146* (9–10), 351–355.

(19) Liang, X.; Sperling, B. A.; Calizo, I.; Cheng, G.; Hacker, C. A.; Zhang, Q.; Obeng, Y.; Yan, K.; Peng, H.; Li, Q.; et al. Toward Clean and Crackless Transfer of Graphene. *ACS Nano* **2011**, *5* (11), 9144–9153.

(20) Dean, C. R.; Young, A. F.; Meric, I.; Lee, C.; Wang, L.; Sorgenfrei, S.; Watanabe, K.; Taniguchi, T.; Kim, P.; Shepard, K. L.; et al. Boron Nitride Substrates for High-Quality Graphene Electronics. *Nat. Nanotechnol.* **2010**, *5* (October), 722–726.

(21) Jang, A.-R.; Hong, S.; Hyun, C.; Yoon, S. I.; Kim, G.; Jeong, H. Y.; Shin, T. J.; Park, S. O.; Wong, K.; Kwak, S. K.; et al. Wafer-Scale and Wrinkle-Free Epitaxial Growth of Single-Orientated Multilayer Hexagonal Boron Nitride on Sapphire. *Nano Lett.* **2016**, *16* (5), 3360–3366.

(22) Couto, N. J. G.; Costanzo, D.; Engels, S.; Ki, D. K.; Watanabe, K.; Taniguchi, T.; Stampfer, C.; Guinea, F.; Morpurgo, A. F. Random Strain Fluctuations as Dominant Disorder Source for High-Quality on-Substrate Graphene Devices. *Phys. Rev. X* **2014**, *4* (4), 1–13.

(23) Neumann, C.; Reichardt, S.; Venezuela, P.; Drögeler, M.; Banszerus, L.; Schmitz, M.; Watanabe, K.; Taniguchi, T.; Mauri, F.; Beschoten, B.; et al. Raman Spectroscopy as Probe of Nanometre-Scale Strain Variations in Graphene. *Nat. Commun.* **2015**, *6* (May), 8429.

(24) Kim, S.; Nah, J.; Jo, I.; Shahrjerdi, D.; Colombo, L.; Yao, Z.; Tutuc, E.; Banerjee, S. K. Realization of a High Mobility Dual-Gated Graphene Field-Effect Transistor with $Al_2O_3$ Dielectric. *Appl. Phys. Lett.* **2009**, *94* (6), 1–4.

# Deterministic patterned growth of high-mobility large-crystal graphene: a path towards wafer scale integration

*Supplementary information*

**Semi-dry transfer**

Initially, copper foils with seeded graphene crystals were coated with a thin support layer of PMMA and left to dry in ambient conditions. A semi-rigid frame made using Kapton adhesive tape (3M) was then attached around the perimeter of the sample. The PMMA membrane with the array of graphene crystals was detached from the Cu substrate using electrochemical delamination using 1M NaOH as electrolyte. The electrochemical reaction was performed by applying a voltage of 2.4V between the sample and a Cu counter electrode (figure S1(a)).

A freestanding PMMA/graphene membrane, stretched out by the Kapton support frame, was removed from the electrolyte, rinsed in water and dried in ambient conditions (figure S1(b)). It was then attached to a micromanipulator stage and aligned to the substrate (figure S1(c)). The membrane was brought into close contact with the target substrate, which was then heated to 100 °C in order to improve the adhesion of graphene. Finally, the PMMA support film was removed by immersing the sample in acetone.

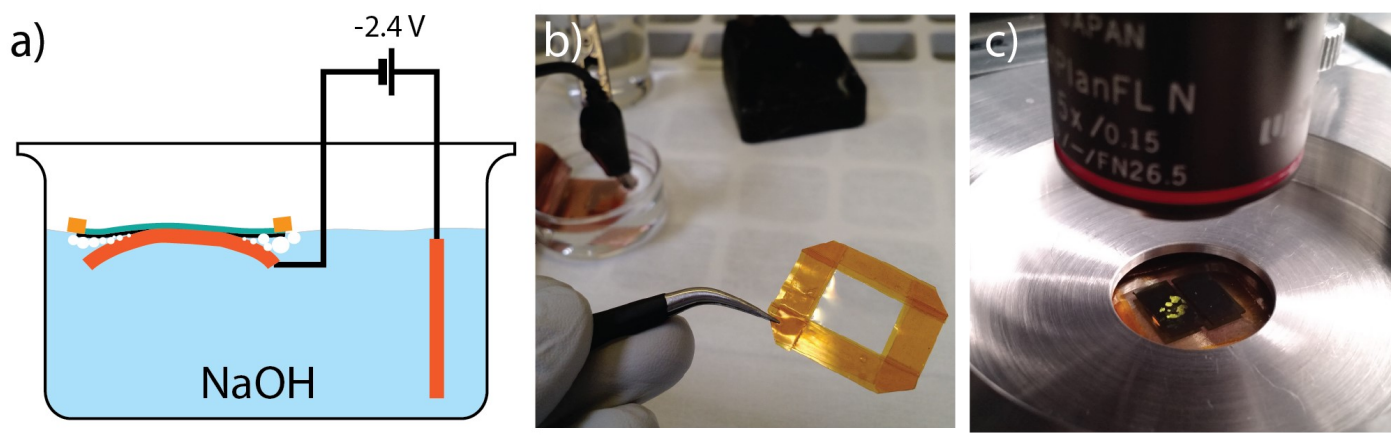

**Figure S1.** Outline of the semi-dry transfer procedure. a) Removal of graphene from Cu growth substrate via electrochemical delamination. b) dry-handling of the graphene array on a suspended PMMA membrane. c) Deterministic placement of the graphene on the target substrate using a mask aligner.

**Raman D-peak mapping**

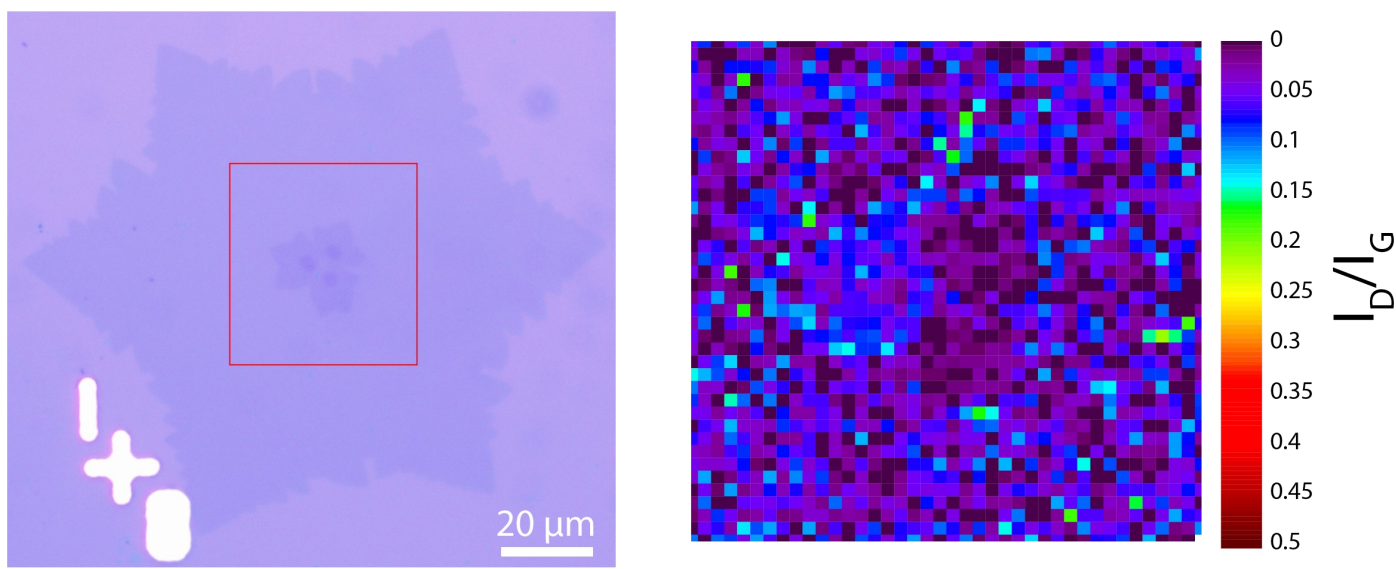

**Figure S2.** Raman D-peak mapping at the centre of a seeded graphene crystal.

**Growth of large-spacing arrays**

Figure S3(a) shows that non-seeded nucleation is less than 10% for arrays with periodicity up to 200 μm. When growing arrays with larger periodicity, namely 500 μm (Fig S3(b)) and even 1 mm (fig. S3 (c)), non-specific nucleation is relatively higher. Although additional crystals are synthesised, Cr seeds are clearly still effective at initiating the growth of large crystals at intended locations.

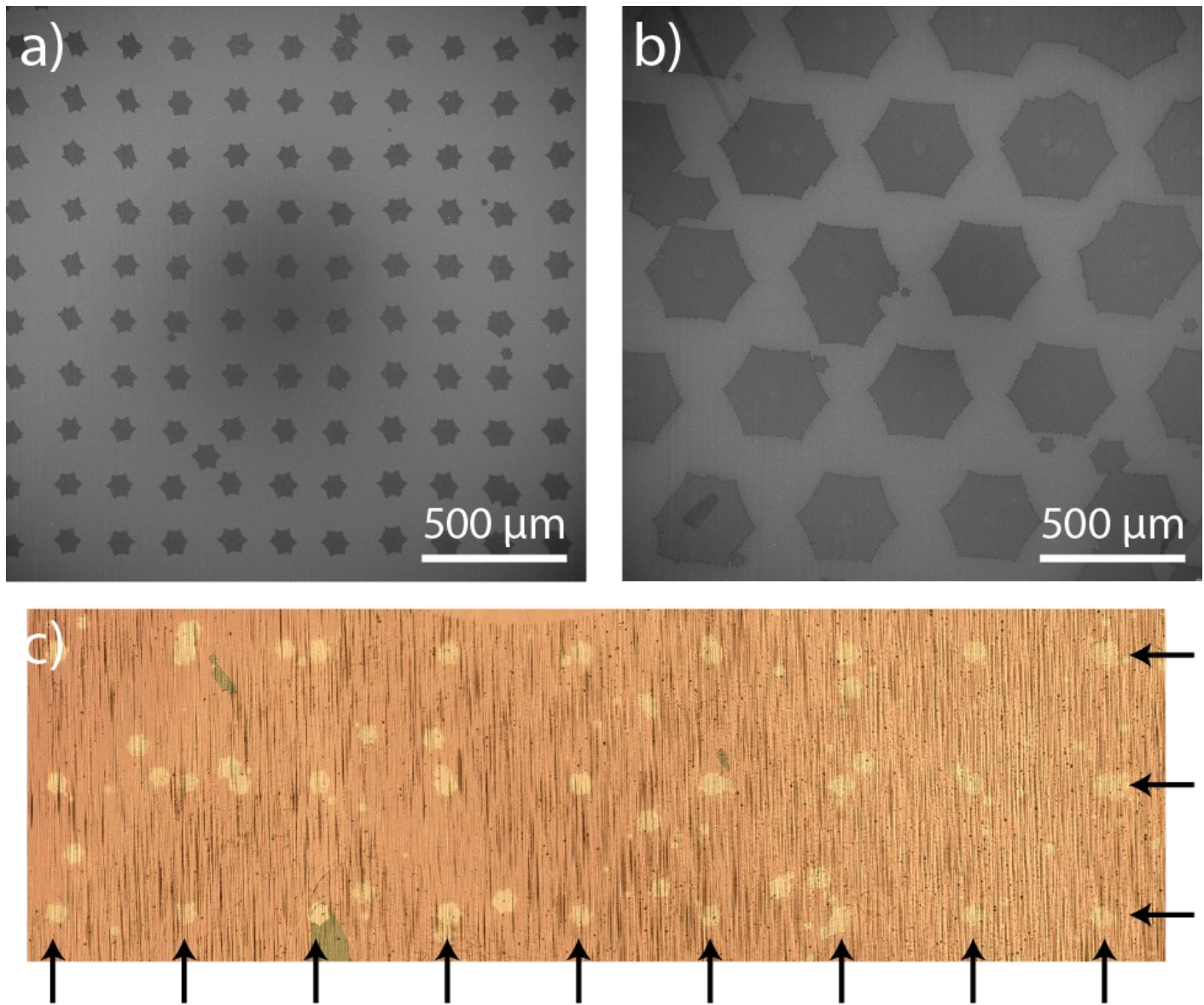

**Figure S3.** Seeded graphene arrays with different periodicities. a) Square array with 200 μm periodicity and 100 μm crystals.